\documentclass[11pt,preprint]{aastex}

\usepackage{graphicx}

\newcommand{\beq}{\begin{equation}}
\newcommand{\eeq}{\end{equation}}
\newcommand{\beqa}{\begin{eqnarray}}
\newcommand{\eeqa}{\end{eqnarray}}

\newcommand{\xv}{{\bf x}}
\newcommand{\rv}{{\bf r}}
\newcommand{\vv}{{\bf v}}
\newcommand{\gv}{{\bf g}}

\def\la{\lower.5ex\hbox{$\; \buildrel < \over \sim \;$}}
\def\ga{\lower.5ex\hbox{$\; \buildrel > \over \sim \;$}}

\begin{document}

\title{Tracing the Magellanic Clouds Back in Time}

\author{P.~J.~E. Peebles}  
\affil{Joseph Henry Laboratories, Princeton University, Princeton, NJ 08544, USA}

\begin{abstract}
A solution is presented for the past motions of the Magellanic Clouds, the Milky Way galaxy, and the Andromeda Nebula, fitted to the measured velocities of the Clouds and M31, under some simplifying assumptions. The galaxies are modeled as isolated bodies back to  redshift about 10, when their velocities relative to the general expansion of the universe were small, consistent with the gravitational instability picture for the growth of structure. Mass outside the Local Group is modeled as a third massive dynamical actor that is responsible for the angular momentum of the Clouds.  A plausible solution under these assumptions requires that the circular velocity of the Milky Way is in the range $200\la v_c\la 230$ km s$^{-1}$.  The solution  seems to be unique up to the modest variations allowed by the choices of $v_c$ and the position of the exterior mass. In this solution the proto-Magellanic Clouds  at high redshift  were near the South pole of the Milky Way (in its present orientation), at physical distance about 200 kpc from the Milky Way and moving away at about 200 km~s$^{-1}$. 
\end{abstract}
\maketitle

\section{Introduction}\label{sec:sec1}

The discovery by Kallivayalil {\it et al.} (2006a,b, hereafter K06a, K06b) that the velocities of the Large and Small Magellanic Clouds (the LMC and SMC) are not much smaller than escape velocity from the Milky Way (MW) opens the possibility of computing where the LMC and SMC were a Hubble time ago. This paper adds to previous studies (Besla {\it et al.} 2007, 2009; Shattow  \& Loeb 2009; Kallivayalil {\it et al.} 2009) an initial condition from cosmology, that at high redshift the motions of the protogalaxies were closer  than now to the Hubble flow. This expresses the well-tested proposition that mass clustering --- and peculiar motions --- grew by gravitational attraction out of small primeval departures from homogeneity. The application of this initial condition is simplified by two major approximations: the protogalaxies are modeled as separate --- though not necessarily condensed --- bodies back to redshift $z_{\rm init}\sim 10$, and the influence of matter outside the Local Group (the LG) is modeled as the effect of a single appropriately placed neighbor with mass comparable to that of M31.

The hypothetical neighbor is introduced to reduce the study of the history of the Magellanic Clouds to a workable problem. The K06a,b velocity measurements place an exceedingly demanding constraint on its solution. The discovery that there is a plausible solution argues for the model, and may be a step toward the more challenging, and perhaps workable, problem of determining whether the observed mass concentrations outside the LG could have been arranged at high redshift so as to act as the third massive body. 

Even with all the simplifying assumptions the solution required the lengthy numerical approach outlined in Section~\ref{sec:sec2}.  I do not imagine this method will find applications to problems outside the Local Group and its more immediate neighbors, but I believe the opportunity to explore in unusually close detail what happened in our immediate extragalactic neighborhood justifies the tangled procedure. 

The main focus of this paper is the analysis of the past history of the LMC in Section~\ref{sec:sec3}. In this computation the mass of the LMC is a parameter to be adjusted along with the other masses to fit to the measured velocities of the LMC and M31. That leads to an LMC mass of about $4\times 10^{10}M_\odot$. K06b point out that at this mass, and within the uncertainty of the measured relative velocity of the Clouds, the SMC can be bound to the LMC, the reasonable situation given their similar positions and motions. Section~\ref{sec:sec4} presents examples of the consistency of this situation in the solution for the motion of the LMC.  Section~\ref{sec:sec5} reviews the arguments for plausibility of the history of the Clouds obtained here.

\section{The Model}\label{sec:sec1a}

The simplifying approximations in this calculation require some comments. The first assumption is that, back to redshift $z_{\rm init}\sim 10$, the momentum and center of mass of a protogalaxy are usefully approximated by the momentum and position of a single mass tracer, in an $N$-body problem with small $N$. This need not be inconsistent with the hierarchical growth of galaxies by merging of substructures: each body is meant to represent the mean motion of the collection of substructures that in the course of time merge to form the galaxy. In the solution obtained here the physical separations of the protogalaxies at $z_{\rm init}\sim 10$ are comparable to their present massive halo radii, meaning the condition for the mass tracer model is that at $z_{\rm init}$ the bulk of each protogalaxy already is within its present halo radius. In particular, the LMC could be in pieces  at $z_{\rm init}$, provided the pieces are scattered over less than about 100~kpc. The model fails if the LMC forms at $z<10$ by the merger of pieces that come together along quite different paths. But different paths suggest relative velocities well above motions within the LMC, which suggests the pieces are not likely to merge. 

The initial condition in the model is that the peculiar velocities of the protoClouds satisfy
\beq
v_{\rm init} \la 100\hbox{ km~s}^{-1}\quad\hbox{ at redshift }\quad z_{\rm init}\sim 10.
\label{eq:LMC_p_v}
\eeq
This expresses the thought that $v_{\rm init}$ might be expected to be comparable to the motions of clumps of matter that are falling into the host protogalaxy rather than moving away from it. To be more explicit, let us note that the peculiar velocity produced by a neighboring concentration of mass $M=3\times 10^{12}M_\odot$ at distance $R=300$~kpc, typical of the solution presented below, produces speed 
\beq
v\simeq tGM/R^2 \simeq 100\hbox{ km~s}^{-1}\quad\hbox{ at redshift }\quad z_{\rm init}\sim 10, 
\label{eq:pec_vel}
\eeq
where $t\sim 2\times 10^{16}$~s is the age of the universe at  $z_{\rm init}$. A peculiar velocity of this order in clouds falling into the protogalaxy is capable of spinning up the disk, and a peculiar velocity of this order is capable of setting the outgoing Clouds along the paths found below. 

In a very simple $N$-body model for the Magellanic Clouds the LMC  and MW would be the only  important dynamical actors. But in this model gravity drives LMC motion directly toward or away from the MW, which is not observed. A three-body model might consist of the MW,  M31, and the LMC, with the SMC acting as a minor perturbation. This puts the gravitational forces in the plane of the MW-M31-LMC triangle, along with the velocities driven by long-range gravitational interactions of the three bodies.  As Kallivayalil {\it et al.} (2009) discuss this also is unacceptable; the K06a measurement indicates the LMC is moving at 270 km s$^{-1}$ normal to the plane. Another dynamical actor seems to be needed to produce this orbital angular momentum. The galaxy M33, which now is about 200~kpc from M31, seems to be too small for the purpose. The minimal potentially viable model thus seems to require a third massive body to represent the effect of matter outside the LG but close enough to it at high redshift to have perturbed the motions of the LG members. The third body is introduced {\it ad hoc}, but as will be described it is supported by the fact that it allows a solution that fits the measured velocities of the LMC and M31. This discussion continues in Section~\ref{sec:sec5}.

\section{Numerical Method}\label{sec:sec2}

The search for motions of the members of the LG under the cosmological initial condition in equation~(\ref{eq:LMC_p_v})  commences with a random trial choice of the masses of the four bodies and the three components of the present position of the hypothetical third body. These seven parameters are used in the numerical action method (NAM:  Peebles 1995; Peebles, Phelps,  Shaya \& Tully 2001; and references therein) to solve the equations of motion under the mixed boundary conditions that the present positions are given and the primeval peculiar velocities match the growing mode of linear perturbation theory in the approximation described in Section~\ref{sec:NAM}. This is a rapid computation for $n_x=20$ time steps, which is adequate to describe the smoothly varying motions of the three massive bodies. It is not adequate for the rapidly changing velocity of the LMC as it approaches the MW. Doubling the number of time steps does not much improve the precision of the present velocity of the LMC and it considerably slows the NAM computation. But the NAM solution at $n_x=20$ gives a useful first approximation to the initial positions and velocities for a conventional numerical integration forward in time for the equations of motion of the four bodies. (The initial conditions are reasonably  well fixed  because the displacements of the bodies at high redshift scale in proportion to the cosmological expansion parameter $a(t)$, to a good approximation, so a linear interpolation in $a$ is a good approximation despite the small number of time steps.) A conventional  forward numerical integration with $p_x=2000$ time steps takes little computation time and is quite adequate to follow the motion of the LMC. This forward integration from the NAM initial conditions shifts the present position of the LMC relative to the MW, but that usually is remedied by a perturbative adjustment of the initial conditions. If the resulting present velocities of M31 and the LMC look promisingly close to what is measured then adjustments of the free parameters can significantly improve the fit to the measured velocities. This last step is not frequently called for, but the short computation time for the prior steps allows many trials.

\subsection{Equations of Motion}\label{sec:eom}

I assume  a cosmological constant and zero space curvature, and ignore radiation energy density, so the cosmological expansion parameter satisfies
\beq
\left({1\over a} {da\over dt}\right)^2 = H_o^2\left({\Omega_m\over a^3} + 1 - \Omega_m\right), \qquad
H_ot = {2\over 3\sqrt{1 - \Omega_m}}\sinh^{-1}a^{3/2}\sqrt{\Omega_m^{-1} - 1},
\label{eq:Friedmann}
\eeq
where the present value of the expansion parameter is $a_o=1$. I adopt $H_o=70$ km s$^{-1}$ Mpc$^{-1}$ and $\Omega_m= 0.25$. 

The $N$-body equations of motion expressed in physical coordinates are
\beq
{d^2\rv_i\over dt^2} = \gv_i + (1 - \Omega_m)H_o^2\vec r_i.
\label{eq:physicalEOM}
\eeq
The last term is the effect of the cosmological constant. (The center of mass placed at the origin of coordinates is not accelerating.) The contributions to the physical acceleration $\gv_i$ of each body by the gravitational attractions of the other bodies are taken to be the simple inverse square law with the exception of the LMC and MW, where I adopt a rigid truncated limiting isothermal sphere model,
\beq
g_{\rm LMC} = {v_c^2\over r}, \qquad 
g_{\rm MW} = {M_{\rm LMC}\over M_{\rm MW}}{v_c^2\over r}, \qquad
\hbox{for } r < r_1 = {GM_{\rm MW}\over v_c^2}, 
\label{eq:physical_g}
\eeq 
with the usual inverse square law at larger separation $r$. The adaption of this model to the computation of the SMC motion is described in Section~\ref{sec:sec4}. The pure inverse square law for all other pairs allows formal solutions with close passages and unrealistic accelerations, but these cases are readily eliminated.  In acceptable solutions the LMC is close to the MW, and equation~(\ref{eq:physical_g}) applies, only at low redshift. All other pair separations are greater than or comparable to usual estimates of halo radii back to $z_{\rm init}\simeq 10$, where the forward integration commences.  

The crude model in equation~(\ref{eq:physical_g}) for the LMC-MW interaction does preserve momentum. The MW structure is known in a good deal more detail than this, but the simplified MW model is an appropriate match to the simplified four-body picture for the LG and its surroundings. Equation~(\ref{eq:physical_g}) ignores the perturbation to the MW mass distribution by the LMC, as in stellar dynamical drag, but in the numerical results presented here the effect on the velocity of the LMC is only about 3 percent if the LMC has recently fallen close to the MW for the first time.\footnote{In the approximation to dynamical drag in Binney and Temaine (1987, section 7.1) the fractional perturbation to the velocity of the LMC by dynamical friction as it falls into the dark halo of the MW is $\delta v/v\sim 1.3\,GM_{\rm LMC}/rv^2\sim 0.03$, where $M_{\rm LMC}$ scatters around the value $4\times 10^{10}M_\odot$ in the solutions in Section~\ref{sec:sec3}, $v$ is the velocity of the LMC, and $r$ is its distance from the MW.}

\subsection{Numerical Action Method}\label{sec:NAM}

The NAM applied here uses comoving coordinates $\xv_i=\rv_i/a(t)$ that describe motion relative to the general expansion of the universe. In these coordinates the equations (\ref{eq:physicalEOM}) of motion with the expansion rate equation~(\ref{eq:Friedmann}) become 
\beq
{d\over dt}\left(a^2{d\xv_i\over dt} \right) = a\,\delta\gv_i, \qquad
\delta\gv_i = \gv_i + {\Omega_mH_o^2\rv_i\over 2a^3}. 
\label{eq:eom}
\eeq
The acceleration $\delta\gv_i$ is the sum of the physical gravitational acceleration produced by the other bodies, as in equations~(\ref{eq:physicalEOM}) and ~(\ref{eq:physical_g}), and the  ``counter'' term that refers the acceleration to that of a homogeneous universe. The effect of the cosmological constant appears here in the time-dependence of $a(t)$. When the counter term just cancels the gravitational interactions, so $\delta\gv_i$ vanishes for all bodies, the bodies may remain at fixed comoving positions, moving apart in accordance with the general expansion of the universe. This balance is increasingly delicate as the redshift increases because the two terms in the right hand side of equation~(\ref{eq:eom}) scale about as $(1+z)^2$ when the motions approximate the general expansion. 

The initial conditions for the growing departure from this balance are modeled after the growing mode of departure from homogeneity in linear perturbation theory for a continuous pressureless mass distribution, where  $a\,\delta\gv_i$ is constant at high redshift (in matter-dominated expansion).  If $a\,\delta\gv_i$ is constant the growing solution to equation~(\ref{eq:eom}) is
\beq
 ad\xv_i/dt=t\delta\gv_i,
\label{eq:eomi}
\eeq
where the expansion time $t$ is measured from $a=0$. 

In the NAM computation the expansion parameter is uniformly spaced at the values
\beq
a_n = (n - 1/2)/(n_x+1/2),
\eeq
with $1\leq n \leq n_x+1$ and present value $a_{n_x+1}=1$.  The earliest time at which positions are computed is at $n=1$. At the half time step before that the expansion parameter vanishes, $a_{1/2}=0$. In leapfrog approximation the differential equation~(\ref{eq:eom}) is replaced by the set of algebraic equations
\beqa
S^k_{i,n} &=& (a^2\dot a)_{n+1/2}{x^k_{i,n+1}-x^k_{i,n}\over a_{n + 1} - a_n}
-(a^2\dot a)_{n-1/2}{x^k_{i,n}-x^k_{i,n-1}\over a_{n} - a_{n-1} }
\nonumber\\
&-& (t_{n+1/2} - t_{n-1/2}) (a_ng^k_{i,n} 
+ \Omega_mH_o^2 x^k_{i,n}/2a_n)= 0, 
\label{eq:NAM}
\eeqa
for the comoving positions $x^k_{i,n}$ of body numbers $i=1$ to 4, Cartesian components $k=1$ to 3, and time steps $n = 2$ to $n_x$. The initial condition in equation~(\ref{eq:eomi}) is applied at the first time step, in the approximation
\beq
S^k_{i,1} = (a^2\dot a)_{3/2}{x^k_{i,2}-x^k_{i,1}\over a_{2} - a_1}
- t_{3/2} (a_1g^k_{i,1} +
 \Omega_mH_o^2 x^k_{i,1}/ 2a_1)= 0.
 \label{eq:NAMi}
\eeq

Equation~(\ref{eq:NAMi}) allows unrealistically large velocities at $n=3/2$, which in the ideal fluid model would correspond to unrealistically large primeval mass fluctuations, but such cases are readily discarded. (In the computation, NAM solutions with LMC initial peculiar velocities greater than 200~km~s$^{-1}$ are rejected. The final step, parameter adjustment to fit the measured velocities, may move the initial peculiar velocity outside this limit.) Equation~(\ref{eq:NAMi}) is only an approximation to the cosmological initial conditions,  because the four-body model is only an approximation to a fluid, and the effect of the distributed mass within each protogalaxy is not modeled at all. These aspects of a more realistic model are supposed to be represented by the freedom allowed to the peculiar motions of the mass tracers representing the protogalaxies at high redshifts.  

The 12 present positions $x^k_{i,n_x+1}$ in the four-body model are given; the $12n_x$ equations (\ref{eq:NAM}) and~(\ref{eq:NAMi}) are to be solved for the $12n_x$ quantities $x^k_{i,n}$ for $1\leq n\leq n_x$. This is termed a numerical action method because the $S^k_{i,n}$ are derivatives of the leapfrog approximation to the action. (It might be mentioned that the addition of a dynamical drag term would remove the relation to an action, but the relaxation to $S^k_{i,n}=0$ with friction would still produce solutions to the discrete leapfrog approximation to the equations of motion.) The NAM solution is obtained by iterated relaxation of the path of each body in turn. The $x^{k'}_{i,n'}$ are adjusted to reduce the $S^k_{i,n}$ to zero using the inverse of the  $3n_x$ by $3n_x$ matrix of derivatives of the $S^k_{i,n}$ with respect to the  $x^{k'}_{i,n'}$. There is not a unique solution at given mixed boundary conditions; the solution this procedure reaches depends on the starting trial paths. The matrix of derivatives of the $S^k_{i,n}$ is written down and its use described in more detail in Peebles (1995). 

\subsection{Forward Numerical Integration}

In the forward numerical integration of the equations of motion the first approximation to the initial physical position $\rv_i$ and velocity $\vv_i$ of body $i$ at the starting value $a_s$ of the expansion parameter are taken from the NAM solution at a chosen time step $n_b$,
\beq
\rv_{i} = a_s(\xv_{i,n_b + 1} + \xv_{i,n_b})/2, \quad
\vv_{i} = a_s^2 H_s {\xv_{i,n_b+1} - \xv_{i,n_b}\over
     a_{n_b + 1} - a_{n_b} } + H_s\rv_{i}, \quad 
     a_s=(a_{n_b + 1} + a_{n_b})/2.
\eeq
The Hubble parameter $H_s$ at $a_s$  is given by equation~(\ref{eq:Friedmann}).  Since the position $x^k_{i,n}$ at high redshift varies with time in a reasonably close approximation to $\delta x^k_{i,n} \propto a(t)$ these interpolations are useful approximations to the wanted initial conditions. 

The solutions presented here use $n_b=2$ from the $n_x=20$ NAM time steps. That is, the initial conditions are interpolated between $n=2$ and $n=3$, and the forward integration commences at redshift
\beq
1+z_s = 1/a_s = 10.25.
\label{eq:zs}
\eeq
I avoid positions at the first NAM time step, $n=1$, because the treatment of the initial condition in equation~(\ref{eq:eomi})  in the discrete form in equation~(\ref{eq:NAMi}) is a rough approximation. 

The forward integration\footnote{The forward computation numerically integrates the equations of motion expressed in the  physical units of equation~(\ref{eq:physicalEOM}). This began as a comforting check of consistency of the two ways to express the equations of motion and the two methods of solution, and I never found occasion to switch to forward integration of the equations of motion expressed in comoving coordinates.}  from these initial conditions at $p_x=2000$ time steps reasonably well reproduces the present positions of the massive bodies given to the NAM computation. The computed position of the LMC is more in error, typically by a few tens of kiloparsecs. Since that is comparable to the present distance between the LMC and MW it must be corrected, and the smaller perturbation to the path of M31 by the LMC may be corrected too. This is done by adjusting the initial positions.\footnote{Adjusting different combinations of initial positions and velocities to get the given present position would produce orbits that have different decaying terms and the same late-time behavior.} Let $e_a$ be the six Cartesian components of the differences between the computed present galactocentric positions of the LMC and M31 and the given present positions. Write the six components of the initial positions of these two galaxies as  $r_b$. Numerically compute the derivatives $d_{a,b}=\partial e_a/\partial r_b$. Invert this 6 by 6 matrix to get corrections to the initial positions, $\delta r_b=-\beta d^{-1}_{b,c}e_c$. It helps to let the constant $\beta$ be less than unity for a few iterations, after which $\beta=1$ usually quickly drives the present positions from the forward integration to the positions given to NAM to machine precision. This procedure slightly perturbs the present position of the third massive body, but that does not matter because this position was chosen at random in the search for an acceptable solution.

\subsection{Fitting the Model to the Measurements}\label{sec:NAMfitting}

The LG galaxies are assigned present heliocentric positions
\beqa
{\rm M31:}\ \ell &=& 121.174, \ b = -21.573, \ d = 750\hbox{ kpc},\nonumber\\
{\rm LMC:}\ \ell &=& 280.465, \ b = -32.888, \ d = 48\hbox{ kpc},\nonumber\\
{\rm SMC:}\ \ell &=& 302.797, \ b = -44.299, \ d = 64\hbox{ kpc},\nonumber\\
{\rm MW:}\ \ell &=& 0, \ b = 0, \ d = 8.5\hbox{ kpc}.
\label{eq:assigned_positions}
\eeqa
The angular positions are from NED, the LMC distance is from Macri {\it et al.} (2006), the SMC distance from North, Gauderon, \& Royer (2009), and the M31 distance is close to the central value of the measurements of the true distance modulus in the summary in Sanna {\it et al.} (2008). The heliocentric redshift of M31 is set to 
\beq
v_r^\odot = -300\hbox{ km s}^{-1},
\label{eq:M31radialvel}
\eeq
and the transverse velocity is free. The galactocentric velocity of the LMC, from K06a, is
\beq
v_x = -86\pm 12\hbox{ km s}^{-1}, \ 
v_y = -268 + v_c - 220 \pm 11\hbox{ km s}^{-1}, \
v_z = 252\pm 16\hbox{ km s}^{-1},
\label{eq:LMCvelocity}
\eeq
and the velocity of the SMC, from K06b, is
\beq
v_x = -87\pm 48\hbox{ km s}^{-1}, \ 
v_y = -247 + v_c - 220 \pm 42\hbox{ km s}^{-1}, \
v_z = 149\pm 37\hbox{ km s}^{-1}.
\label{eq:SMCvelocity}
\eeq
In a standard convention these are Cartesian velocity components in a right-hand coordinate system where the positive $x$-axis is directed from the Solar System to the center of the MW and the $y$-axis points in the direction of rotation of the Solar System around the MW. Following Shattow \& Loeb (2009) I consider the possibility that the circular velocity of rotation of the MW  may be significantly different from the standard value, $v_c=220$~km~s$^{-1}$.  I ignore the correction to the velocity of M31 for our motion relative to the local standard of rest. 

The computation of the path of the LMC seeks a minimum of the statistic $\chi_4^2$ summed over the four differences between the model and the measured velocity components in equations~(\ref{eq:M31radialvel}) and~(\ref{eq:LMCvelocity}). The measurement uncertainties in equation~(\ref{eq:LMCvelocity}) are treated as standard deviations, and the heliocentric radial velocity of M31 is assigned a nominal standard deviation of 4 km~s$^{-1}$. At given $v_c$ the seven free parameters are  the masses of the four bodies and the present position of the third massive body. To make $\chi_4^2$ vary smoothly with the variation of the parameters (apart from the multi-value effect to be noted next) the numerical values of the position errors $e_a$,  and the values of the $S^k_{i,n}$ that represent the equations of motion, are relaxed to zero to near double precision accuracy. This is much closer to zero than is needed for the comparison of measured and model velocities, but the smooth variation aids the search for minima of $\chi_4^2$.

There are two barriers to the reduction of $\chi_4^2$ to an acceptable value. First, this statistic is a many-valued function of the parameters, and a small change in a parameter can produce a large shift in paths, usually with a jump to quite unacceptable present velocities. Second, the trial adjustment can end at a local minimum of $\chi_4^2$ that is unacceptably large. The remedy in both cases is to start again with different randomly chosen trial parameters and trial orbits for the NAM solution. This random search places the third body uniformly at random in the heliocentric sky at distance $d_3=3 + 3{\cal R}$~Mpc, where the random number ${\cal R}$ is uniformly distributed between 0 and 1. The logarithm of the mass of the MW is uniformly distributed in the factor of two range centered on $1.5\times 10^{12}M_\odot$, the M31 mass in the factor of three range around $2.0\times 10^{12}M_\odot$, the third body mass in the factor of 4 range around $3.0\times 10^{12}M_\odot$, and the LMC mass in the factor of 2 range around $3\times 10^{10}M_\odot$. The parameters may end up outside these ranges as they are adjusted to minimize $\chi_4^2$. An overnight run on a MAC Mini can sample about 8000 random trials. That typically yields two to four promising-looking cases for $\chi_4^2$ minimization at  $v_c=220$ km s$^{-1}$, and about one case at $v_c=250$ km s$^{-1}$. 

\begin{table}[htpb]
\centering
\begin{tabular}{lccccccccc}
\multicolumn{9}{c}{Table 1: Parameters}\\
\tableline\tableline
\vspace{2pt}
  & $M_{\rm MW}^{\rm a}$ & 
$M_{ M31}^{\rm a}$ & $M_3^{\rm a}$ & 
$M_{\rm LMC}^{\rm a}$ & $r_1^{\rm b}$ & {$\ell_3$} & {$b_3$} & 
${d_3}^{\rm c}$ & 
${cz_3}^{\rm d}$\\
\tableline
200a & 12.8 &   31.6 &   29.6 &   0.84 &   138 &   21 &  $-57$ &  4.6 &  260 \\  
200b  & 11.7 &   39.2 &   27.9 &   0.54 &   126 &  173 & 14 &    3.4 &   154\\
220a & 12.6 &   32.9 &   41.6 &   0.55&  112 &  29 &   42 &    6.0 &   431 \\
220a$^\ast$ & 12.6 &   31.1 &   41.1 &   0.42 &  112 &  30 &    42 &    4.1 &   239 \\
220b & 10.3 &   37.0 &   31.0 &   0.21 &  92 & 191 &    13 &    5.4 &   456 \\
230a  & 12.1 &    32.1 &   41.3 &   0.40 &   98  &  29 &    42 &    6.0 &   418 \\
230b & 10.2 &   37.0 &   26.9 &   0.13 &  83 &  191 &     14 &    5.3 &    453 \\
240  &   9.7 &   35.8 &   12.8 &   0.13 &    72 &  24 &    49 &    6.1 &   470\\
250 &    8.8 &   24.9 &   28.4 &   0.58  &  61 &  220 &    81 &    3.3 &   320\\
200c & 14.4 &   13.4 &   96.9 &   0.36 & 154 &  121 &   $-20$ &    5.7 &   214\\
\tableline
\multicolumn{7}{l}{$^{\rm a}10^{11} M_\odot$,\quad $^{\rm b}$kpc, \quad
$^{\rm c}$Mpc, \quad$^{\rm d}$km s$^{-1}$}
\end{tabular}
\end{table}

\section{The Motion of the LMC}\label{sec:sec3}

\subsection{Model Parameters and Solutions}

Table~1 lists values of the adjustable model parameters in solutions with an acceptable present velocity of the LMC and, except for the last solution, an acceptable redshft of M31. The circular velocity $v_c$ of the MW is indicated in the model label in the first column. The next four columns are the model masses.  Next is a derived quantity, the cutoff radius $r_1$ of the mass distribution in the MW in the truncated isothermal model halo in equation~(\ref{eq:physical_g}). The next three columns list the present heliocentric position of the third massive body, in galactic coordinates. The last column is another derived quantity, the present heliocentric  radial velocity $cz_3$ of the third body.

\begin{table}[htpb]
\centering
\begin{tabular}{lrrrrrrrrrrrr}
\multicolumn{13}{c}{Table 2: Present Velocities$^{\rm a}$}\\
\tableline\tableline
\vspace{2pt}
 & \multicolumn{4}{c}{LMC} & &\multicolumn{7}{c}{M31}\\
\cline{2-5} \cline{7-13}\\
\vspace{-30pt}\\
  & {$v_x$} & {$v_y$} & {$v_z$} & {$\chi_3^2$} &{} &  {$v_x$} & {$v_y$} & {$v_z$} & {$v_\ell$} 
& {$v_b$}& {$v_r$} & $v_r^\odot$\\
\tableline
200a & $  -81$ & $ -285 $ & $ 251$ & $0.3$ & & 39  &$-106$ &  75 & $ 22$ & $ 29$ & $ -131$ & $ -290 $ \\
200b & $  -81$ & $ -282 $ & $ 235$ & $1.6$ & & 54 & $-116$ & 61 & $ 14$ & $ 10$ & $ -140$ & $  -300 $\\
220a & $-89$ & $-283$ &  272 &    3.7  & & 45 & $-108$ & 46 & 17 & $1$ & $-125$ & $-300$\\
220a$^\ast$ & $-89$ &  $-282$ &   271 &    3.1 & & 41 & $-106$ & 39 & 20 & $-5$ & $ -119$& $-294$ \\
220b & $-64$ &  $-281$ &   246 &    4.7 & & 37 & $-106$ & 53 & 23& 9 &  $-121$ & $-296$ \\
230a & $-96$ &  $-265$ &   291 &    7.1 & & 42 & $-103$ & 41 & 17 &  $-2$&  $-118$ & $-301$ \\
230b  & $-84$ & $ -264$ & $264$ &  0.9 && 36  &$-103$ &  50 & $    22$ & $     8$ & $  -118$ & $  -301 $ \\
240  &  $-88$ &  $-261$ &   267 &    2.3 & & 48 & $-97 $ & 36 &  9 &    $-6$ &  $-113$ & $-304$\\
250  & $-94$ &  $-244$ &   288 &    5.8 & & 26 & $-79$ & 60 & 18 &    26 &   $-98$ & $-297$\\
200c  & $-85$ &  $-297$ &   257 &    0.7 & & 41 & 8 & 81 & $-39$ &    70 &   $-43$ & $-202$\\
\tableline
\multicolumn{7}{l}{$^{a}$km s$^{-1}$}
\end{tabular}
\end{table}

Present velocities in the model solutions are  tabulated in Table 2.\footnote{It will be recalled that the results presented here and the other figures and tables are from the numerical integration of the equations of motion forward in time. The NAM solution is required to get a useful starting approximation to the initial conditions, but it plays no role in the computation after that.} The first column repeats the model label in Table~1. The next three columns are the Cartesian components of the galactocentric velocity of the LMC (in the coordinate system in eqs~[\ref{eq:LMCvelocity}] and~[\ref{eq:SMCvelocity}]). The values of $\chi_3^2$ in the fifth column are the sums of the three differences of measured and model LMC velocity components, with the measurement uncertainties in equation~(\ref{eq:LMCvelocity}) treated as standard deviations. The next three columns are the Cartesian components of the galactocentric velocity of M31, the next three translate that to the galactocentric transverse and radial velocities of M31, and the last column is the heliocentric radial velocity of M31. 

\begin{table}[htpb]
\centering
\begin{tabular}{lcccccccc}
\multicolumn{9}{c}{Table 3: Initial Conditions$^{\rm a}$}\\
\tableline\tableline
\vspace{2pt}
 & \multicolumn{4}{c}{separations$^{\rm b}$} &  &
  \multicolumn{3}{c}{relative peculiar velocities$^{\rm c}$}\\
  \cline{2-5} \cline{7-9}\\
\vspace{-25pt}\\ 
& LMC-MW & LMC-M31 & M31-MW & 3-LG & & LMC-MW & LMC-M31 & M31-MW\\
\tableline
200a &   217 &   401 &   499 &   219 &&   225 &   267 &   336 \\
200b &   186 &   404 &   486 &   256 &&    60 &   281 &   282\\
220a &   177&  267&  345 & 414 &&   87&  117&   83 \\
220a$^\ast$ &    179 &   245 &   328 &   366 &&    87 &    90 &    44\\
220b & 183&  281&  355 & 377 &&   95&   81&  116 \\
230a & 175 &  256 &  335 & 398 &&   93 &  105 &   72 \\
230b  &   181 &   279 &   351 &   356 &&    97 &    80 &   114\\
240 &  162&  284&  316 & 440 &&   93&  129&   66\\
250 & 183 &  235 &  301 & 417 &&  325 &  111 &  307\\
200c & 213 &  406 &  285 & 427 &&   91 &  108 &   78  \\
\tableline
\multicolumn{7}{l}{$^{\rm a}$at $1+z_{\rm init}=10.25$\quad $^{\rm b}$physical lengths, kpc\quad 
$^{\rm c}$km s$^{-1}$}
\end{tabular}
\end{table}

All models in Tables 1 and 2 are reasonable fits to the measured LMC velocity. All except the last fit the measured redshift of M31, within reasonable allowance for the crude nature of the four-body model. The discrepancy in the redshift of M31 in the last model is beyond reasonable allowance. This case illustrates the existence of at least one solution to the four-body problem that fits the measured velocity of the LMC and gives M31 a transverse velocity comparable to its radial galactocentric velocity.  The large transverse velocity of M31 could be reconciled with its measured radial velocity by increasing the masses of M31 and the MW, but attempts in that direction spoil the fit to the velocity of the LMC.

\subsection{Initial Conditions and the Role of the Third Massive Body}

Table~3 lists initial conditions at $z_{\rm init}\sim 10$. Columns two to four are the initial physical separations of the LG galaxies, next is the initial physical distance of the third massive body from the center of mass of the LG, and the last three columns are the magnitudes of the initial relative peculiar velocities ($\vv_p=\vv - H_s\rv$, where $\rv$ and $\vv$ are the physical relative position and relative velocity of the pair of protogalaxies and $H_s$ is the value of Hubble's constant at the start of the integration).

\begin{figure}[htpb]
\begin{center}
\includegraphics[angle=0,width=4.in]{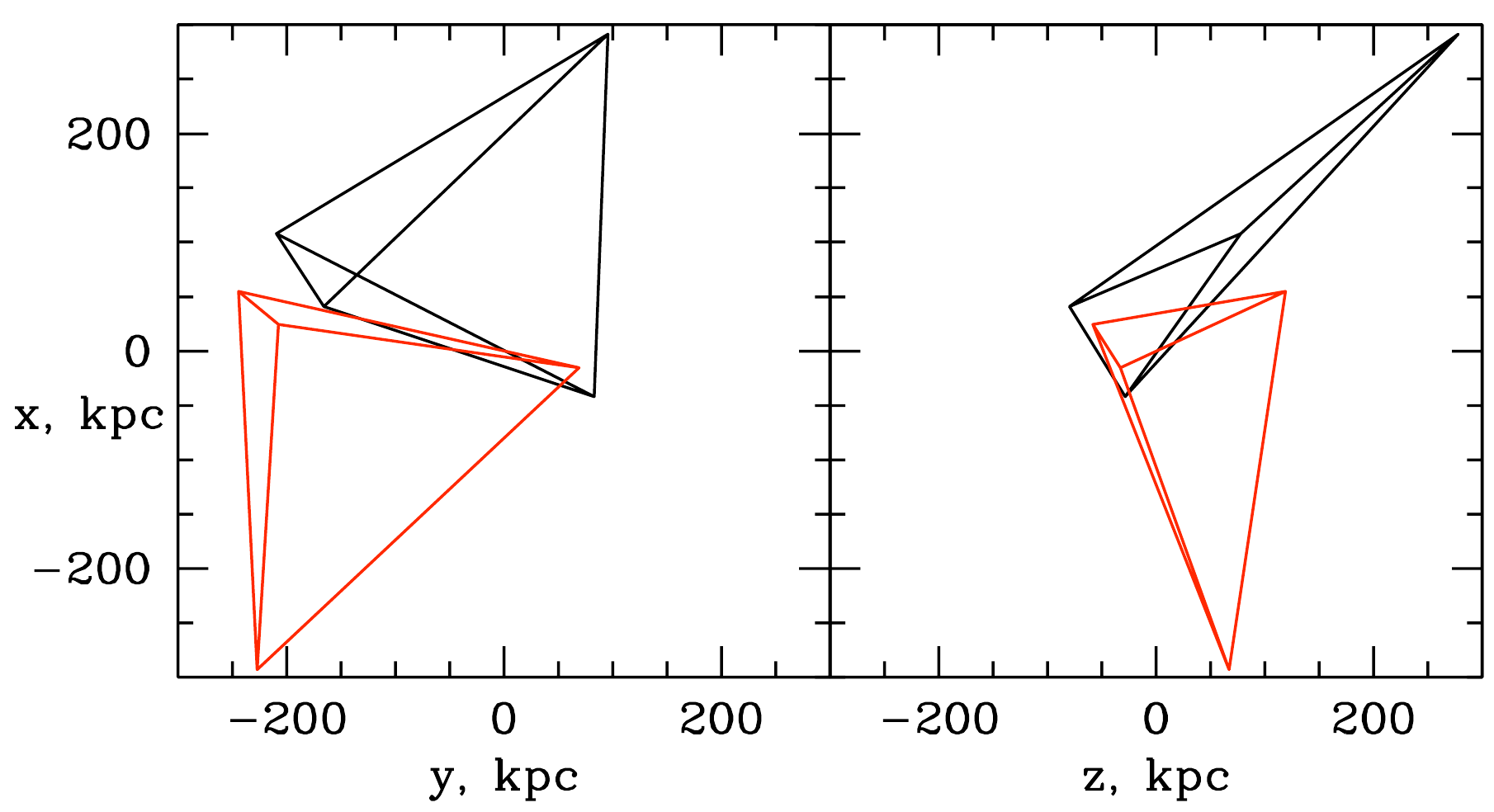} 
\caption{\small Initial positions of the LG galaxies for solution 220a (black) and 220b (red). The third massive body is near the top of the figure in 220a and near the bottom in 220b. In both solutions M31 is the lowest of the LG galaxies, MW is highest and the LMC is in the middle.\label{figure:1}}
\end{center}
\end{figure}

At given $v_c$ different solutions have similar LG masses and initial relative positions and velocities, while the angular position of the third massive body can be quite different (Table~1). To understand this recall that small initial peculiar velocities require small departures from the balance of gravitational acceleration with the cosmological counter term (eq.~[\ref{eq:eom}]), while a departure from exact balance is required to allow the tidal field of the third  body to produce the present orbital angular momentum of the LG. Figure~\ref{figure:1} illustrates the two ways this can happen. The initial positions in solutions 220a and 220b are at the vertices of the black and red tetrahedrons, with the origins of the coordinates set so the center of mass of the LG is at the origin. The similar sizes of the LG triangles in the two solutions are largely set by the similar masses, with some perturbation by the tidal field of the third body. The height and tilt of the tetrahedrons relative to the LG bases are set by the near balance of gravity and the counter term, and by the departure from balance that torques the LG. This can happen on either side of the LG plane, in the paired solutions 200a and 200b, 220a and 220b, and 230a and 230b. (I have not looked for paired solutions at larger $v_c$.) The tetrahedrons in these paired solutions are not mirror images, because that would produce oppositely directed angular momentum of the LG. The tetrahedrons on opposite sides of the LG in Figure~\ref{figure:1} are tilted in the different way required to give the LG the same angular momentum about the M31-MW axis.

Two degeneracies are suggested by these considerations. In the first, illustrated by solutions 220a and 220a$^\ast$, the present distance to the third body is substantially larger in 220a. That is because the third body had larger initial velocity away from the LG. The present tidal field on the LG is stronger in 220a$^\ast$, but that does not much matter because in these solutions the tidal field at low redshift has little effect. The freedom to adjust the present distance of the third body has not been explored in other solutions.

A second degeneracy that has not been explored at all is the relation between the third body mass $M_3$ and its initial distance $r_3$ from the LG. The condition of near balance of gravity and the counter term for the third body is approximately $G(M_{\rm LG}+M_3)/r_3^2 \sim G\rho r_3$, where $\rho$ is the mean mass density in the background homogeneous cosmology. The tidal field on the LG scales as 
\beq
{\rm tidal~force}\sim  M_3/r_3^3 \sim\rho\,  M_3/(M_{\rm LG}+M_3).
\eeq
If the third body is more massive than the LG the dynamical model is expected to allow the mass of the third body to scale with its initial distance as 
\beq
M_3\propto r_3^3.
\label{eq:m3_scaling}
\eeq
This may figure in the assessment of how the third massive body in this analysis might be related to the present distribution of mass around the LG.

\begin{figure}[htpb]
\begin{center}
\includegraphics[angle=0,width=6.5in]{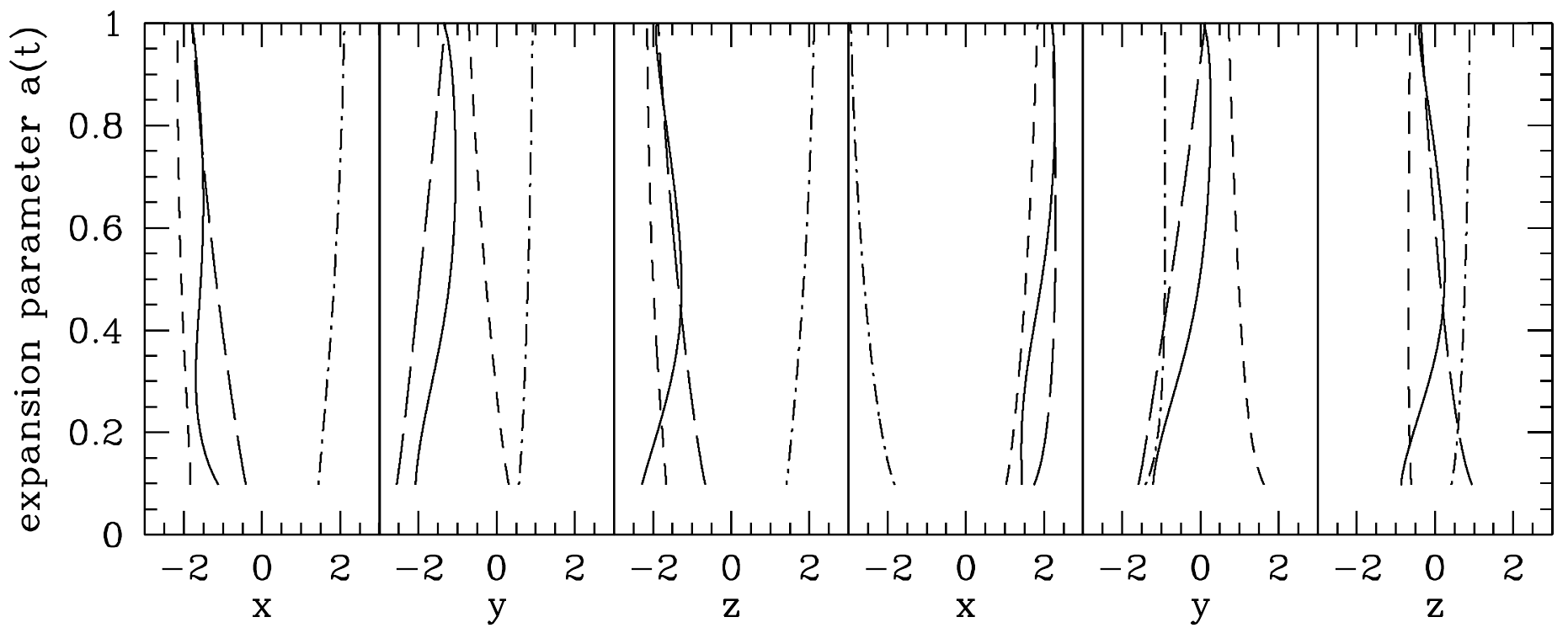} 
\caption{\small Orbits in solutions 220a (left) and 220b (right) in comoving coordinates. The  path of the LMC is plotted as the solid curve, the MW as the long dashed curve, M31 as the short dashed curve, and the third massive body as the dot-dashed curve.}
\label{figure:2}
\end{center}
\end{figure}

\subsection{The Pattern of Motions Within the Local Group}

Figure \ref{figure:2} compares orbits in solutions 220a and 220b with the initial conditions shown in Figure~\ref{figure:1}. The comoving coordinates are centered on the center of mass of the four-body system and oriented as in equation~(\ref{eq:LMCvelocity}). The  distinct difference between the solutions is seen for example in the position of the massive third body, plotted as the dot dash-line, well to the right of the LG in the $x$-direction in 220a, and well to the left in 220b (in the first and fourth panels in the figure). But one may also notice the similarity of relative motions within the LG.

\begin{figure}[htpb]
\begin{center} 
$
\begin{array}{cc}\hspace{-12pt}
\includegraphics[angle=0,width=3.7in]{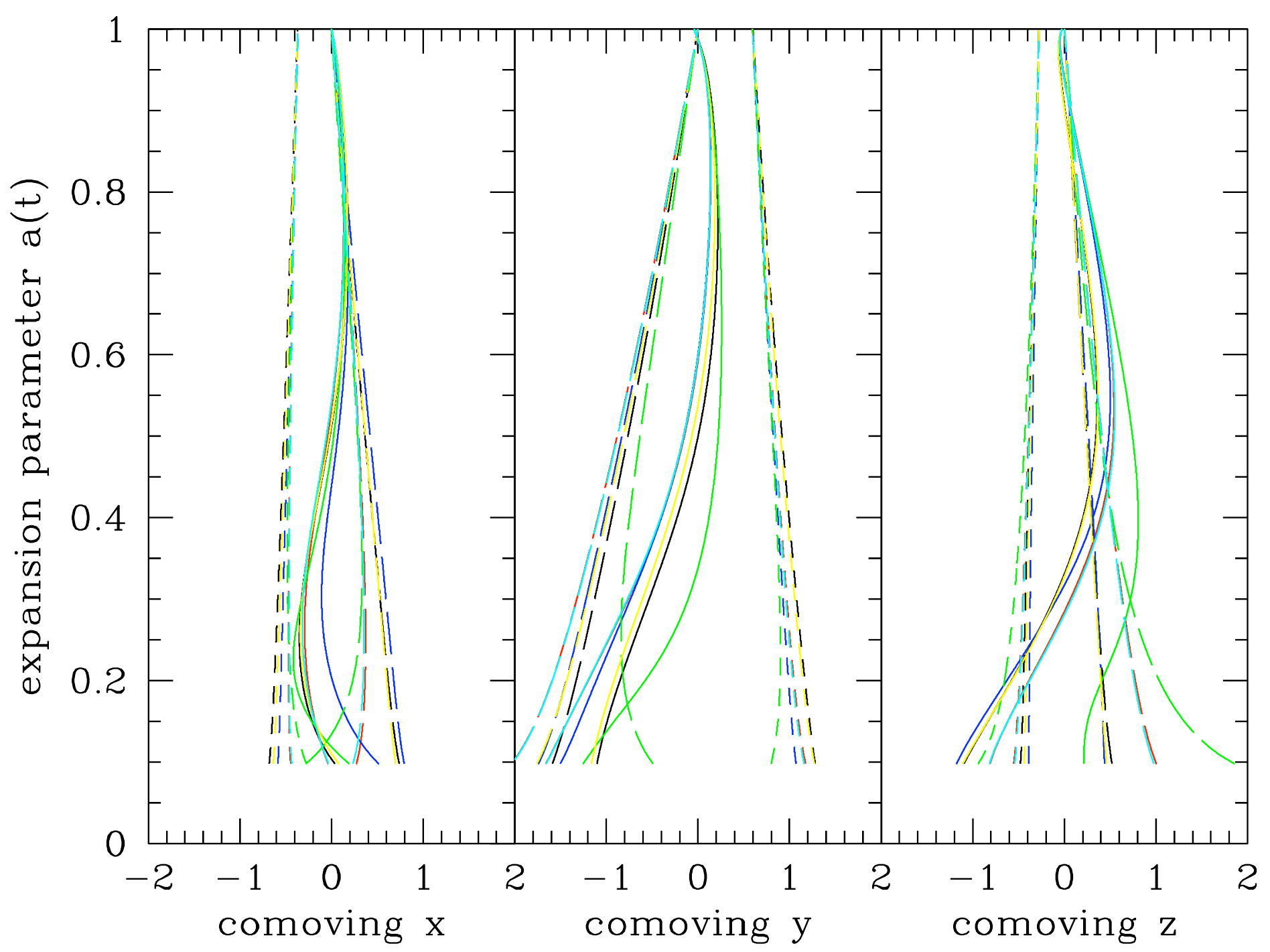} & \hspace{2pt}
\includegraphics[angle=0,width=2.8in]{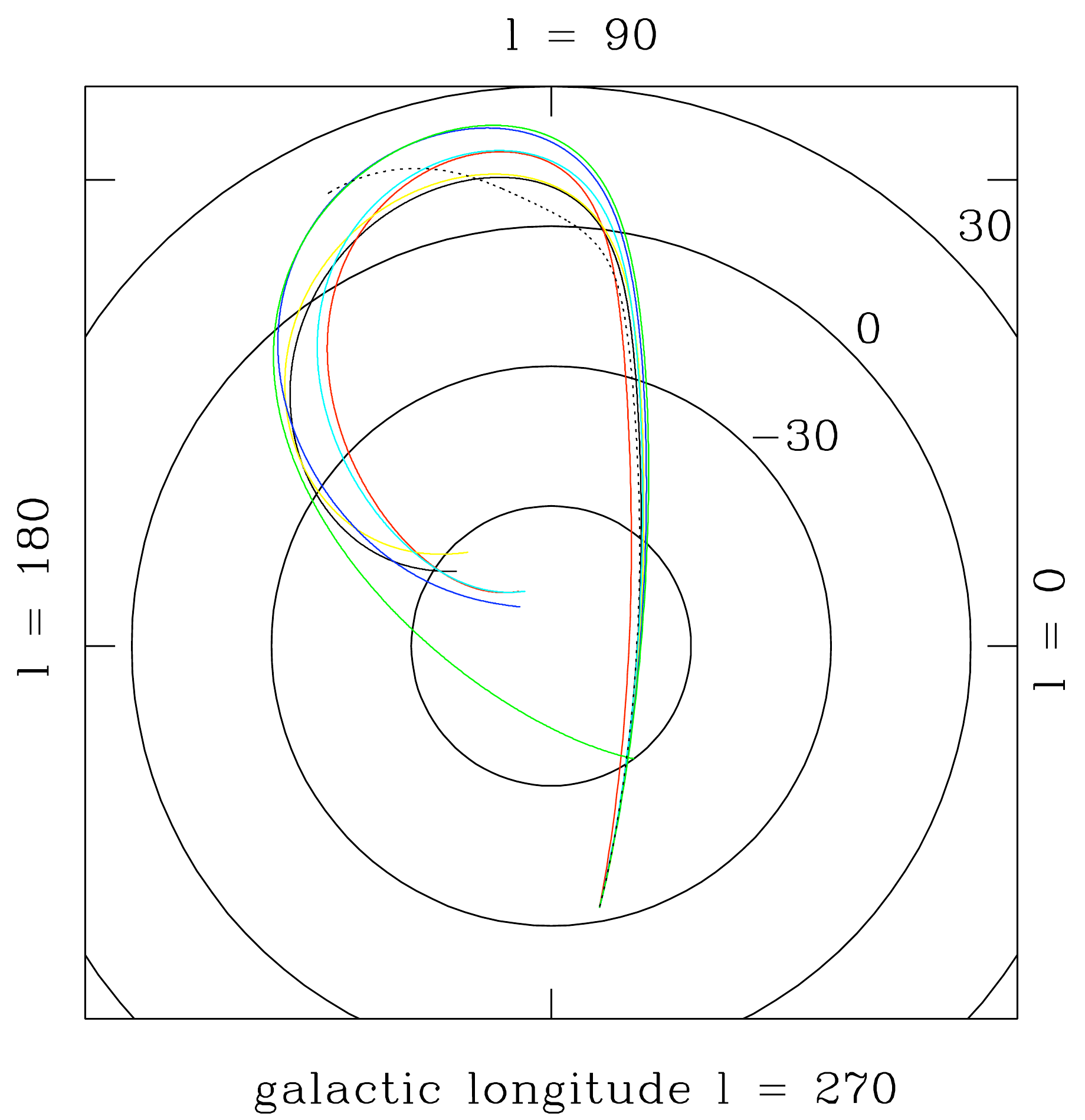} 
\end{array}
$
\caption{\small Relative motions of the Local Group galaxies in solution 220a (black), 220b (red), 230a (yellow), 230b (cyan), 240 (blue), 250 (green), and, in the right-hand panel, 200c (dotted black). On the left the galaxy identifications are the same as in Figure~\ref{figure:2}, but the orbits are plotted relative to the center of mass of the LG with an offset to put the present position of the MW at the origin. On the right are paths of the LMC in heliocentric galactic coordinates.\label{figure:3}}
\end{center}
\end{figure}

The similarity of solutions within the LG is more clearly seen in Figure~\ref{figure:3}. In the panel on the left the comoving coordinates are referred to the center of mass of the three LG galaxies, but with the offset that puts the present position of the MW at the origin. The panel on the right shows paths of the heliocentric angular position of the LMC in galactic coordinates. The paths in both panels certainly differ, and the differences increase with increasing redshift, but there is a distinct pattern around which the solutions, including 200a and 200b, scatter. 

As in the polar plot for the solution in Figure~19 in Besla {\it et al.} (2007), the LMC traces back in time to galactic latitude $b\simeq -70^\circ$ as it swings pass $\ell =0$.  At still higher redshift the paths in the LG model solutions swing toward larger galactic longitude, and, apart from 200c, then trace back to initial positions near the south pole of the MW (in its present orientation). Solution 200c is most different, though its path does bend to larger longitude at higher redshift.
 
All the acceptable solutions I have found share the pattern of relative motions of the LG galaxies in Figure~\ref{figure:3}. This is the basis for the conclusion that under the hypothesis of a single important external dynamical actor a solution for the relative motions of the LG galaxies consistent with the velocities of the LMC and M31 exists and has a unique pattern. 

\section{Motion of the SMC}\label{sec:sec4}

\begin{figure}[htpb]
\begin{center}
\includegraphics[angle=0,width=4.5in]{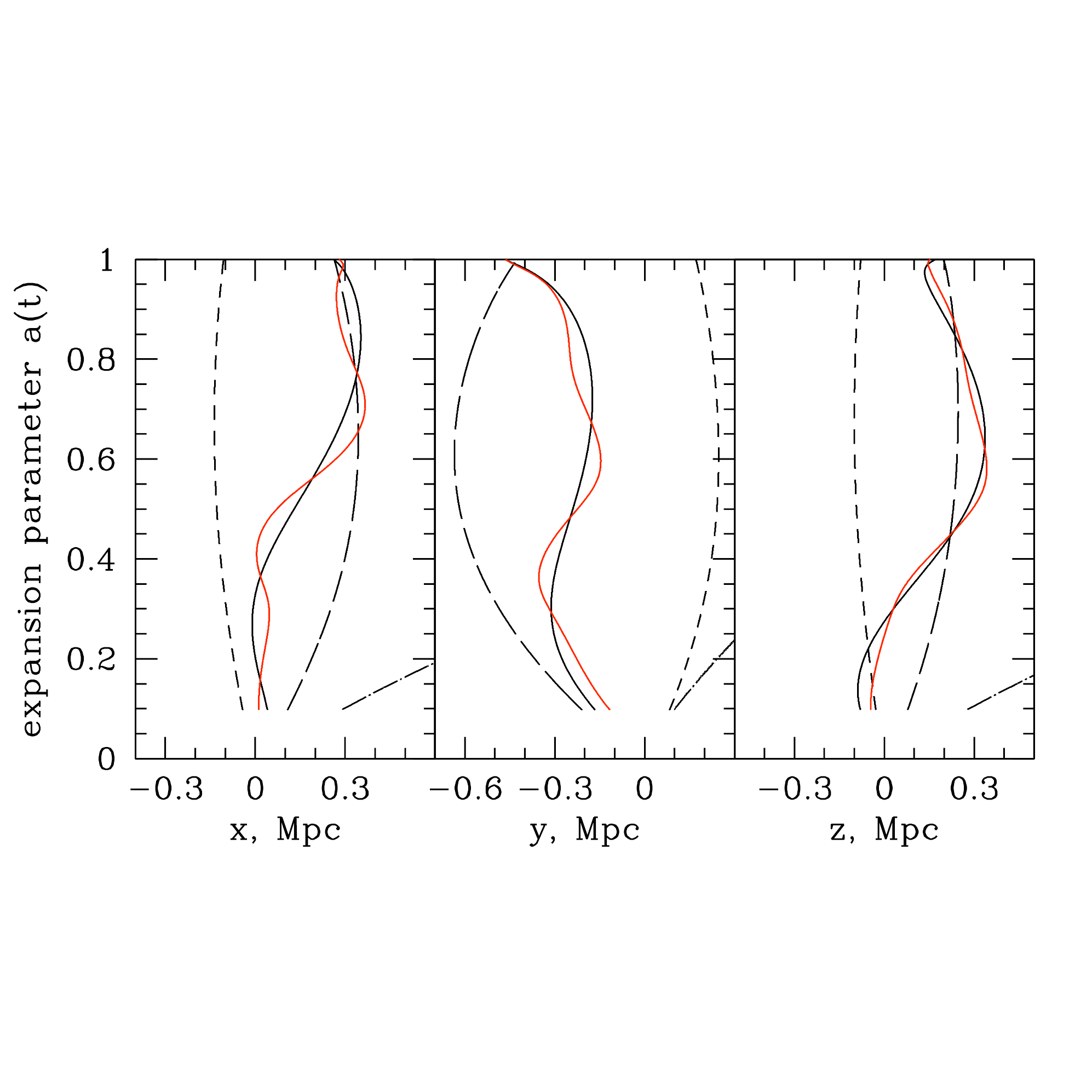} 
\caption{\small Motions relative to the center of mass of the LG in solution 220a, as in Figures~\ref{figure:2} and~\ref{figure:3}, except here plotted in physical coordinates. The red curve is the path of the SMC in example (a).\label{figure:4}}
\end{center}
\end{figure}

Figure~\ref{figure:4} shows an example of an orbit for the SMC that fits the K06b SMC velocity measurement. The SMC is assigned zero mass --- it is a test particle --- so it can be added to any of the solutions discussed in the last section. This example uses solution 220a.

For the purpose of computing the SMC motion the LMC is assigned a rigid spherical mass distribution with density run $\rho\propto r^{-2}$ inside a sharp cutoff at radius $r_l$, as in the model in equation~(\ref{eq:physical_g}). The adopted circular velocity and cutoff radius for the LMC are $v_c=100$ km~s$^{-1}$ and $r_l= 24$~kpc, for the LMC mass in solution 220a. (In the computation of the motions of the bodies that have mass the LMC is treated as a point particle in the rigid MW halo. The replacement with a distributed mass in the LMC for the purpose of computing the path of the SMC is not a serious inconsistency at the hoped-for accuracy of the  four-body model.) The SMC path is from numerical integration of the equations of motion of the test particle back in time with the same 2000 time steps used in the forward numerical integration used to follow the path of the LMC. The present conditions for this integration back in time are the SMC position in equation~(\ref{eq:assigned_positions}) and, in the first trial, the central value of the measured  SMC velocity (eq.~[\ref{eq:SMCvelocity}]). The present velocity is than adjusted, by trial, to get an orbit in which the SMC avoids the more massive galaxies and is reasonably close to the position and velocity of the LMC at $z_{\rm init}\sim 10$. 

\begin{figure}[htpb]
\begin{center}
\includegraphics[angle=0,width=3.in]{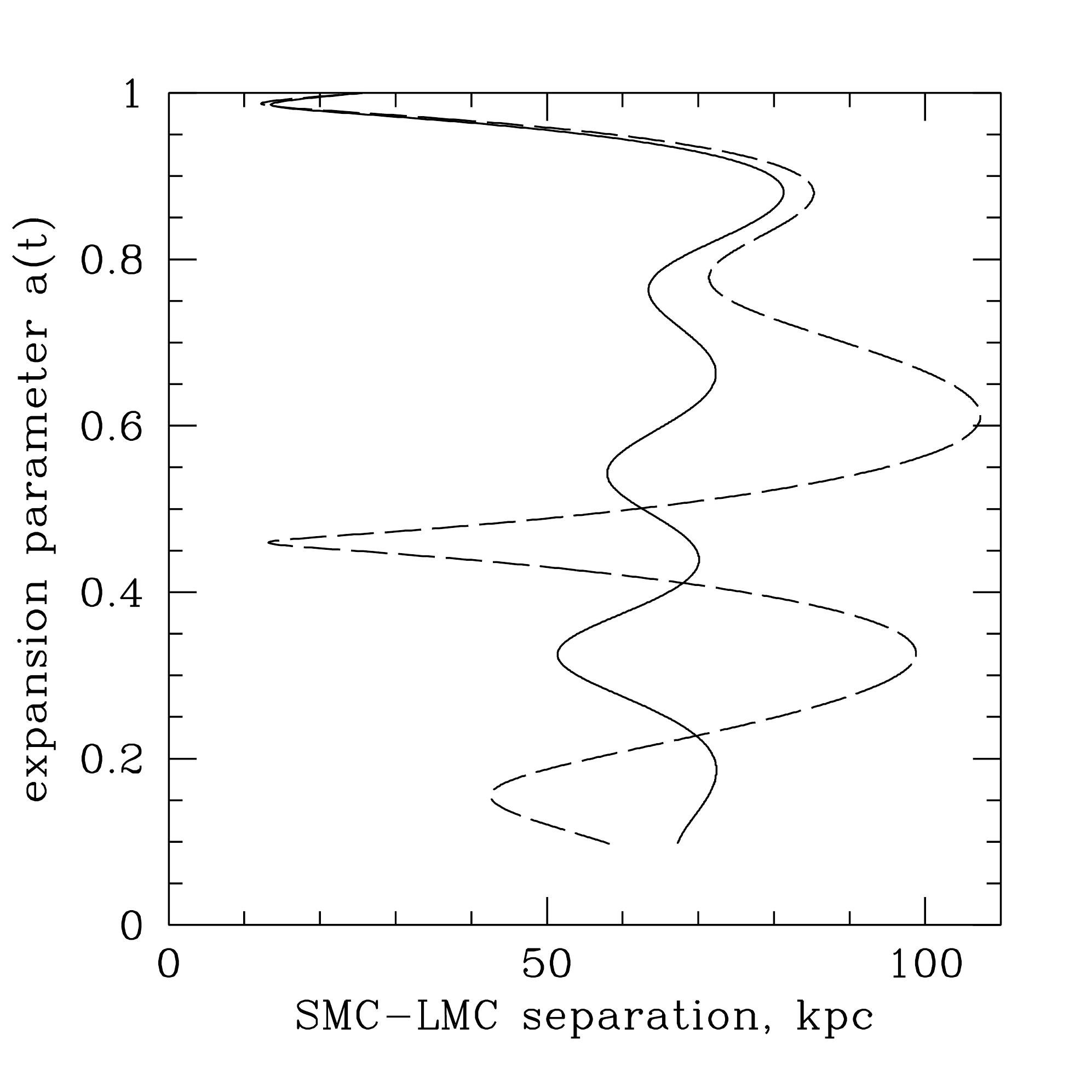} 
\caption{\small Physical separation of the SMC and LMC plotted as the solid curve in example (a) (eq.~[\ref{eq:SMCmodelvelocity_a}]) and the dashed curve in example (b) (eq.~[\ref{eq:SMCmodelvelocity_b}]).\label{figure:5}}
\end{center}
\end{figure}

The present velocity of the SMC in example (a) in Figure~\ref{figure:4} is 
\beq
v_a = -107,\ -250,\ 148\hbox{ km s}^{-1},
\label{eq:SMCmodelvelocity_a}
\eeq
well within the uncertainty of the measurement (eq.~[\ref{eq:SMCvelocity}]). At $z_{\rm init}\sim 10$ the physical separation and relative velocity of the Clouds are
\beq
r_a = 67\hbox{ kpc}, \qquad v_a = 61\hbox{ km s}^{-1}.
\label{eq:SMCmodel_a}
\eeq
Figure \ref{figure:5} shows the evolution of the physical separation of the Clouds in this example and in a second example (b) that has present  velocity
\beq
v_b= -91,\ -242,\ 148\hbox{ km s}^{-1},
\label{eq:SMCmodelvelocity_b}
\eeq
also consistent with the measurement, and initial separation and relative velocity
\beq
r_b = 58\hbox{ kpc}, \qquad v_b = 66\hbox{ km s}^{-1}.
\label{eq:SMCmodel_b}
\eeq
In both examples there is a recent close passage, 200~Myr ago at minimum separation $r_{\rm mn}=14$~kpc and relative velocity $v_{\rm mn}=165$~km~s$^{-1}$ in (a), and 180~Myr ago at $r_{\rm mn}=12$~kpc and $v_{\rm mn}=170$~km~s$^{-1}$ in (b). This is quite similar to the most recent close passage of the Clouds in the solutions in K06b, Figure 13, and Besla {\it et al.} (2009), Figure~ 2. In the LG model the Clouds in example (a) have been orbiting each other at separation 50 to 70~kpc prior to that, while in (b) there is another close approach at redshift $z\sim 1$. The case for the recent close passage seems reasonably good. What happened before that is not well constrained by the dynamical model. 

\section{Discussion}\label{sec:sec5}

The starting assumptions in this analysis are that the motions of the matter now concentrated around the LMC, M31 and the MW may be adequately represented by the paths of mass tracers; the peculiar velocities of the mass tracers at redshift $z_{\rm init}\sim 10$ are small, consistent with the gravitational instability picture for structure formation; the influence of matter outside the Local Group may be adequately represented by the gravitational field of an appropriately placed third massive body; and the Magellanic Clouds have returned to the MW for the first time since moving away from it at high redshift. Several considerations in addition to those mentioned in Section~\ref{sec:sec1a} bear on these assumptions and on the credibility of the resulting dynamical model for the Local Group.  

It is worth emphasizing again that the mass tracer model in this analysis does not require that the galaxies had their present compact structures at high redshift. It does require that merging histories since $z_{\rm init}\sim 10$ have been local  enough that the momentum and center of mass of the pieces of a protogalaxy are usefully represented by a single mass tracer. We have an example in the motions of the Magellanic Clouds in Figure~\ref{figure:4}. If in the future the Clouds merged a future analysis of this sort  would model the  Clouds as a single body. Tracing that situation back in time would miss the earlier presence two mass concentrations, but it would give a reasonable indication of where the matter in the merged galaxy came from. 

\begin{table}[htpb]
\centering
\begin{tabular}{lrrrrrr}
\multicolumn{7}{c}{Table 4: Comparison to Linear Theory$^{\rm a}$}\\
\tableline\tableline
\vspace{2pt}
 & \multicolumn{2}{c}{$v_x^{\rm b}$} & \multicolumn{2}{c}{$v_y$} & \multicolumn{2}{c}{$v_z$} \\
\tableline
MW &$-32$&$-15$ &     29 &   9  &$-35$& $-15$ \\
M31 &  9   & $-20$&$ -36 $&$-8$& $-3$ & $-19$ \\
3   & 4    & 20     &   19 &      4 &  12    & 19  \\
LMC &$-90$&$-15$&      17 &   22   &   29   &  60 \\
\tableline
\multicolumn{7}{l}{$^{\rm a}$at $1+z_{\rm init}=10.25$\quad $^{\rm b}$km s$^{-1}$}
\end{tabular}
\end{table}

The second of the starting assumptions is that the peculiar velocities of the mass tracers at $z_{\rm init}\sim 10$ are consistent with what would be produced by the gravitational interactions with neighbors and, in the case of the Clouds, with the higher multipoles of the mass distributions within the nearest massive protogalaxies. A measure of the former situation is presented in equation~(\ref{eq:pec_vel}). A more direct measure is shown in Table 4. Under the header for each Cartesian velocity component, the first column lists the initial values of the peculiar velocity components for each LG protogalaxy in numerical solution 220a. The second column under each header is the prediction from linear perturbation theory applied at $z_{\rm init}$ to a continuous pressureless fluid, $\vv = t\delta\gv$ (eq.~[\ref{eq:eomi}]), where the peculiar gravitational acceleration $\delta\gv_i$ of body $i$ is  computed from the positions at $z_{\rm init}$ in solution 220a. 

The numerical and perturbation theory velocity components in Table~4 are correlated, though with considerable scatter. The scatter may be in part an effect of the nonlinear growth of clustering, but almost certainly a large contributing factor is that a four-body system is not a very good approximation to a continuous fluid. But the important point for our purpose is that the initial peculiar velocities in the numerical solutions are not out of line with what would be produced by the gravitational interactions among the protogalaxies supposed to have been present at redshift $z_{\rm init}\sim 10$. Also worth noting is that the LMC at $z_{\rm init}$  is moving about as fast as matter within protogalaxies, including what is needed to spin up disks. This likely is about as far as one can take considerations of naturalness of initial conditions within this schematic four-body model. 

Before considering the third major assumption, the representation of the mass external to the LG, let us review results of the model based on this assumption. Because there are seven free parameters in Table~1 (in addition to the choice of the circular velocity $v_c$) to fit four constraints --- the four measured velocity components singled out for particular attention --- one might have expected that the parameters could be adjusted to match the central values of the velocity measurements. That is not guaranteed, of course; there could have been no parameter choice that yields acceptable  velocities. Thus it should be counted as a positive result that the dynamical model can fit the velocities, as indicated in Table~2. It may also be significant that the allowed paths of the LG galaxies seem to follow the unique pattern shown in Figure~\ref{figure:3}.  

Another arguably positive result is that the small transverse motion of M31 in the dynamical model fitted to the redshift of M31 and the velocity of the LMC agrees with the quite independent line of argument by van der Marel \& Guhathakurta (2008). They find that the galactocentric velocity of M31 is 
\beq
v_x=97 \pm 35,\ v_y=-67\pm 26,\ v_z=80\pm 32 \hbox{ km s}^{-1}.
\label{eq:M31_velocity}
\eeq
This may be compared to the velocities in columns 6 to 8 in Table~2. In solution 220a the velocity components differ from equation~(\ref{eq:M31_velocity}) by 1.5, 1.6, and 1.1 times the stated uncertainties. Within the limitations of these two very different ways to estimate the velocity of M31 I consider this to be encouraging consistency. The situation is similar for the other solutions apart from 200c, where $v_y$ differs from the van der Marel \& Guhathakurta result by three times the stated error, and the radial velocity of M31 is unacceptable. 

Allowing the mass of the LMC to be a free parameter aids the search for acceptable matches to the measured velocites. The LMC masses found this way, and listed in Table 1, scatter around $4\times 10^{10}M_\odot$. This may be biased by the choice of the central value in the range of LMC masses  in the search for parameters that produce promising NAM solutions. Apart from the parameter adjustment following the NAM search, which allows masses outside the search range, I have not tried to find acceptable solutions with larger LMC masses. But it is reasonable to count as another positive result the existence of minima of the measure $\chi_4^2$ of goodness of fit at LMC masses close to the range, 1 to $3\times 10^{10}M_\odot$, that K06b considered, and at a mass that Besla (private communication) finds reasonable from astrophysical considerations. At $M_{\rm LMC}=4\times 10^{10}M_\odot$ the mass-to-light ratio of the LMC is $M/L_B\sim 20$ Solar units. In solution 220a the mass of M31 is $3.3\times 10^{12}M_\odot$, which makes its mass-to-light ratio $M/L_B=110$. The difference seems to be larger than could be reasonably accounted for by the younger star population in the LMC. Perhaps the LMC lost dark matter to its nearest neighbor at high redshift, the MW, leaving the LMC with a relatively low mass-to-light ratio. Perhaps the search for acceptable solutions in the dynamical model should be extended to larger LMC masses.

It is reasonable to suppose the SMC has been close to the LMC in the past, and K06b point out that, within the uncertainty in the relative velocity of the Clouds, conventional estimates of the mass of the LMC can make this so. The examples in Figures~\ref{figure:4} and~\ref{figure:5} of SMC orbits bound to the LMC  illustrate this, and show that at $z_{\rm init}\sim 10$ the Clouds could have been on the outskirts of the protoMW and moving away from it along similar paths. This would make the Clouds among the last of the larger debris to fall back toward the MW. Notable in these examples, and in the orbits in K06b and Besla {\it et al.} (2009), is that the Clouds passed close to each other a few hundred million years ago. The reproducibility --- though based on the same constraints on the relative velocity --- suggests this close passage may really have happened. The relative velocity at minimum separation, $\sim 170$ km~s$^{-1}$, is large enough, and the separation, about 15~kpc, small enough that one might imagine the Clouds escaped merging but bear observable marks of the encounter. The  dynamical model cannot give significant guidance to what happened to the Clouds prior to this recent close encounter. The large gas content might be taken to suggest that example (a) in Figure~\ref{figure:5}, in which the Clouds were not closer than 50~kpc prior to the close encounter, is more plausible than example (b), with its earlier close encounter. The existence of the former solution for the SMC might be counted as additional though not very strong evidence for the dynamical model. 

If the circular velocity of the MW is in the range 
\beq
200\la v_c\la 230\hbox{ km s}^{-1},\label{eq:acceptable_vc} 
\eeq
then the model requires that the mass of the MW is 30\% to 40\% of the mass of M31. That is reasonably commensurate with the larger circular velocity of M31, about $250$~km~s$^{-1}$ (Carignan {\it et al.} 2006). However, if the MW circular velocity is  larger,  $v_c=240$ to 250~km~s$^{-1}$, the model requires that the mass of M31 is about three times that of the MW, while the circular velocities of the two galaxies are similar, a questionable situation. 

Ghez {\it et al.} (2008) find that their measurement of the distance to the center of the MW with the  Reid \& Brunthaler (2004) proper motion of Sagittarius A* indicates $v_c=229\pm 18$ km~s$^{-1}$. The Reid {\it et al.} (2009) measurements of distances and proper motions of star-forming regions in the MW fitted to a model for the rotation curve indicate $v_c=254\pm 16$~km~s$^{-1}$. If the circular velocity proved to be 230~km~s$^{-1}$,  which is within the Ghez {\it et al.} measurement and outside Reid {\it et al.} by 1.5 times the uncertainty, it would be a comfortable fit to the dynamical model. If, on the other hand, the circular velocity proved to be 250~km~s$^{-1}$, which is  within the Reid {\it et al.} uncertainty and outside Ghez {\it et al.} by 1.2 times the uncertainty, there would be three interpretations to consider. First, following  Shattow \& Loeb (2009), the LMC may have completed several orbits around its host, the MW. In this case, tracing the LMC back to the situation at high redshift likely is too complicated for the dynamical methods used here. But arguing against this interpretation is van den Bergh's (2006) point: if the Clouds have already completed several close passages of the MW why is their gas content anomalously high for a satellite this close to its host? Second, the net dark masses of M31 and the MW may not be closely related to the circular velocities in the flat parts of their rotation curves. There would seem to have been room for fortuitous exchange of largely dark matter between the outskirts of the protoMW and protoM31 at $z\sim 10$, which could have made the MW considerably less massive than M31 even if the inner circular velocities are similar. Third, the four-body model may be a useful approximation but not accurate enough to merit close attention to such details as the difference between the apparently acceptable situation at $v_c=230$ km~s$^{-1}$ and the apparently questionable situation of the model at $v_c=250$ km~s$^{-1}$. Perhaps the model prediction in equation~(\ref{eq:acceptable_vc}) is as close to reality as one could have hoped for. 

A major open issue of accuracy is the third of the starting  assumptions, the treatment of the mass outside the LG in this dynamical model. The external mass seems to be needed to push the motion of the LMC out of the plane defined by the MW, M31 and the LMC. It is modeled as a single body with mass comparable to that of M31. Perhaps this is a realistic picture. Perhaps more realistic would be a larger mass at greater initial distance, scaled as in equation~(\ref{eq:m3_scaling}). Or perhaps the wanted tidal field on the LG is produced by several different external mass concentrations. The present positions of the nearest major masses outside the LG are known. The challenge is to analyze where these masses might have been at high redshift, and to determine whether their positions then could have produced the tidal field indicated by the dynamical model for the LG. Previous experience with the considerable variety of solutions to the problem of tracing the motions of galaxies back in time, together with the tangled method of computation of motions within the LG, suggest that addressing this issue will be complicated. The evidence that the transverse velocity of M31 is small is encouraging, however, for it would considerably reduce the multiplicity of solutions illustrated in Figures 6 to 9 in Peebles {\it et al.} (2001). Addressing this issue might best be done in steps, perhaps beginning with another problem we can be reasonably sure is workable, the exploration of the allowed range of masses and initial and present distances of the third body. 

The methods of analysis in this paper have a  semiempirical character that complement what is learned from large-scale numerical simulations of structure formation in the $\Lambda$CDM cosmology, as in  Li and  White (2008). Further comparisons of what both approaches indicate likely will continue to be instructive. On the phenomenological side, it may be timely to reconsider the analysis by Loeb {\it et al.} (2005) of the orbit and survival of the stellar disk of M33, and also the orbit and survival of the H{\small I} disk of IC10, using the Brunthaler {\it et al.} (2005, 2007) proper motions and the dynamical model of the Local Group presented here. Measurements of the positions and velocities of the more isolated dwarf members of the Local Group, including  NGC~6822, may be particularly useful because their orbits seem likely to be simple enough to be reasonably well constrained within the model, though that is to be checked.

The central theme of this paper is that the measurements of the proper motions of the Magellanic Clouds by Kallivayalil {\it et al.} (2006a,b), together with the initial conditions suggested to us by cosmology, have greatly tightened constraints on the dynamics of the Local Group. Further progress in proper motion measurements, as in the Gaia and SIM projects, and, equally important, continued advances in the art of distance measurements, will further tighten the constraints, and may show us whether we really understand how dynamics is operating in our immediate extragalactic neighborhood.

\acknowledgments 

I am grateful to Gurtina Besla, Nitya Kallivayalil, Avi Loeb, Adi Nusser, Steven Phelps, Ed Shaya, Brent Tully, and Roeland van der Marel for their stimulating discussions and collaborations in developing methods of analysis, and to {\it Numerical Receipes} (Press {\it et al.} 1992) for the matrix inversions.


\begin{thebibliography}{}

\bibitem[Besla et al.(2007)]{2007ApJ...668..949B} Besla, G., Kallivayalil, 
N., Hernquist, L., Robertson, B., Cox, T.~J., van der Marel, R.~P., 
\& Alcock, C.\ 2007, \apj, 668, 949

\bibitem[Besla et al.(2009)]{2009IAUS..256...99B} Besla, G., Kallivayalil, 
N., Hernquist, L., van der Marel, R.~P., Cox, T.~J., Robertson, B., 
\& Alcock, C.\ 2009, IAU Symposium, 256, 99

\bibitem[Binney \& Tremaine(1987)]{1987gady.book.....B} Binney, J., \& Tremaine, S.\ 1987, Princeton, NJ, Princeton University Press

\bibitem[Brunthaler et al.(2005)]{2005Sci...307.1440B} Brunthaler, A., 
Reid, M.~J., Falcke, H., Greenhill, L.~J., \& Henkel, C.\ 2005, Science, 307, 1440 

\bibitem[Brunthaler et al.(2007)]{2007A&A...462..101B} Brunthaler, A., Reid, M.~J., Falcke, H., Henkel, C., \& Menten, K.~M.\ 2007, \aap, 462, 101

\bibitem[Carignan et al.(2006)]{2006ApJ...641L.109C} Carignan, C., Chemin, 
L., Huchtmeier, W.~K., \& Lockman, F.~J.\ 2006, \apjl, 641, L109 

\bibitem[Ghez et al.(2008)]{2008ApJ...689.1044G} Ghez, A.~M., et al.\ 2008, 
\apj, 689, 1044 

\bibitem[Kallivayalil et al.(2006)]{2006ApJ...638..772K} Kallivayalil, N., 
van der Marel, R.~P., Alcock, C., Axelrod, T., Cook, K.~H., Drake, A.~J., 
\& Geha, M.\ 2006a, \apj, 638, 772 (K06a)

\bibitem[Kallivayalil et al.(2006b)]{2006ApJ...652.1213K} Kallivayalil, N., 
van der Marel, R.~P., \& Alcock, C.\ 2006b, \apj, 652, 1213  (K06b)

\bibitem[Kallivayalil et al.(2009)]{2009arXiv0905.4283K} Kallivayalil, N., 
Besla, G., Sanderson, R., \& Alcock, C.\ 2009, arXiv:0905.4283

\bibitem[Li\& White(2008)]{2008MNRAS.384.1459L} Li, Y.-S., \& White, S.~D.~M.\ 2008, \mnras, 384, 1459 

\bibitem[Loeb et al.(2005)]{2005ApJ...633..894L} Loeb, A., Reid, M.~J., 
Brunthaler, A., \& Falcke, H.\ 2005, \apj, 633, 894 

\bibitem[Macri et al.(2006)]{2006ApJ...652.1133M} Macri, L.~M., Stanek, 
K.~Z., Bersier, D., Greenhill, L.~J., \& Reid, M.~J.\ 2006, \apj, 652, 1133 

\bibitem[North et al.(2009)]{2009IAUS..256...57N} North, P.~L., Gauderon, 
R., \& Royer, F.\ 2009, IAU Symposium, 256, 57

\bibitem[Peebles(1995)]{1995ApJ...449...52P} Peebles, P.~J.~E.\ 1995, \apj, 449, 52

\bibitem[Peebles et al.(2001)]{2001ApJ...554..104P} Peebles, P.~J.~E., 
Phelps, S.~D., Shaya, E.~J., \& Tully, R.~B.\ 2001, \apj, 554, 104 

\bibitem[Press et al.(1992)]{1992nrfa.book.....P} Press, W.~H., Teukolsky, 
S.~A., Vetterling, W.~T., \& Flannery, B.~P.\ 1992, Cambridge: University Press, 2nd ed. 

\bibitem[Reid \& Brunthaler(2004)]{2004ApJ...616..872R} Reid, M.~J., \& Brunthaler, A.\ 2004, \apj, 616, 872

\bibitem[Reid et al.(2009)]{2009ApJ...700..137R} Reid, M.~J., et al.\ 2009, 
\apj, 700, 137 

\bibitem[Sanna et al.(2008)]{2008ApJ...688L..69S} Sanna, N., et al.\ 2008, 
\apjl, 688, L69 

\bibitem[Shattow \& Loeb(2009)]{2009MNRAS.392L..21S} Shattow, G., \& Loeb, A.\ 2009, \mnras, 392, L21 

\bibitem[van den Bergh(2006)]{2006AJ....132.1571V} van den Bergh, S.\ 2006, 
\aj, 132, 1571 

\bibitem[van der Marel \& Guhathakurta(2008)]{2008ApJ...678..187V} van der Marel, R.~P., \& Guhathakurta, P.\ 2008, \apj, 678, 187

\end{thebibliography}
\end{document}